\documentclass[11pt,a4paper]{article}
\usepackage{jheppub,amsmath,  amssymb,slashed,url,bm,textgreek,upgreek,braket}
\usepackage{graphicx}
\usepackage{epstopdf}
\def\t{{ \sf t}} 
\def\tt{{\mathfrak t}}

\def\a{a}

\def\be{\begin{equation}}
\def\ee{\end{equation}}

\def\h{\widehat}

\def\O{{\mathcal O}}

\def\A{{\mathcal A}}

\def\[{\bigl [}

\def\]{\bigr ]}

\def\N{{\mathcal N}}

\def\L{{\mathcal  L}}
\def\t{\widetilde }
\def\h{\widehat}

\def\H{{\mathcal H}}

\def\bar{\overline}
\def\bA{{\bar A}}

\font\teneurm=eurm10 \font\seveneurm=eurm7  \font\fiveeurm=eurm5
\newfam\eurmfam
\textfont\eurmfam=\teneurm \scriptfont\eurmfam=\seveneurm
\scriptscriptfont\eurmfam=\fiveeurm

\font\teneusm=eusm10 \font\seveneusm=eusm7 \font\fiveeusm=eusm5
\newfam\eusmfam
\textfont\eusmfam=\teneusm \scriptfont\eusmfam=\seveneusm
\scriptscriptfont\eusmfam=\fiveeusm

\font\tencmmib=cmmib10 \skewchar\tencmmib='177
\font\sevencmmib=cmmib7 \skewchar\sevencmmib='177
\font\fivecmmib=cmmib5 \skewchar\fivecmmib='177
\newfam\cmmibfam
\textfont\cmmibfam=\tencmmib \scriptfont\cmmibfam=\sevencmmib
\scriptscriptfont\cmmibfam=\fivecmmib

\def\Tr{{\rm{Tr}}}
\def\PA{{\rm{PA}}}
\def\i{{\rm i}}
\def\E{{\mathbb E}}
\def\sw{{\sigma_W}}
\def\eps{\epsilon}
\def\rL{{\sigma_{{\mathrm L}}}}
\def\cE{{\mathcal E}}
\def\E{{\mathbb E}}
\DeclareMathOperator{\csch}{csch}

\title{A Note on Corrections to Entanglement Wedge Reconstruction}

 \author{Edward Witten}
\affiliation{School of Natural Sciences, Institute for Advanced Study,\\ 1 Einstein Drive, Princeton, NJ 08540 USA}
\abstract{If entanglement wedge reconstruction is exact, then (under certain assumptions)  the area term in the RT formula is a $c$-number, indicating that the choice of a bulk quantum state does not influence  the geometry.   Recently  Cao, Cheng, Karthikeyan, Li, and Preskill considered a generic perturbation away from exact entanglement wedge reconstruction.  The optimal reconstruction was defined; based on this, an effective area function that depends nontrivially on the quantum state was defined and its properties were analyzed.   Here we make one aspect of this picture more quantitative, by showing that if as expected the area term in the RT formula is of order $1/G$ while the bulk entropy is of order 1, then the corrections to entanglement wedge reconstruction are exponentially small (in $G$) relative to corrections to the area function.   In the framework under discussion, there is an area function but no area operator; we discuss to what extent this is the expected behavior in holography.
}

\begin{document}\maketitle

\section{Introduction}\label{intro} 

Entanglement wedge reconstruction has been an important idea in understanding holographic duality between quantum gravity in the bulk of spacetime and ordinary quantum field theory on the boundary \cite{Ent1,Ent2,Ent3,Ent4,Ent5,Ent6}.   However, it is known that  if entanglement wedge reconstruction is  exact then (under certain assumptions, discussed shortly) the area term in the Ryu-Takayanagi (RT) formula is a constant, independent of the quantum state of the bulk fields  \cite{Harlow}.   Differently 
put, if entanglement wedge reconstruction is exact, then there is no gravitational backreaction of bulk matter on the area of the RT surface.  Since this would be true for the RT surface of any boundary region, one would expect that actually there is no gravitational backreaction of matter  on the geometry at all in such a scenario.

Therefore, it is expected that in a realistic model of quantum holography,
entanglement wedge reconstruction is not exact. Departures  from exact entanglement wedge reconstruction have
been discussed by several authors     \cite{Kelly, Ent6,Corr2,Corr3,Corr4,Corr5}.   Recently, Cao, Cheng, Karthikeyan, Li, and Preskill (CCKLP) considered a generic small perturbation
away from exact entanglement wedge reconstruction \cite{CCKLP}.    In  a context in which exact  boundary reconstruction of the bulk state is not possible, they defined
the optimal reconstruction.   Based on this, they  simply defined  the area   term in the RT formula as  
 the difference between the entropy of the full state in the fundamental description and the entropy of the optimally reconstructed bulk state.    They further investigated  some  properties of this area function.   Notably, this area function is not the expectation in the quantum state of an area operator.\footnote{The CCKLP approach is certainly not the only proposal in 
 the literature with an area function on the set of quantum states but no area operator.   See \cite{AkersPenington} for a quite different alternative.}
 
 In this article, we make one aspect of the CCKLP setup more quantitative.   We show that if,  as expected, 
 the area term in the RT formula is of order $1/G$ (where $G$ is Newton's constant), while the bulk term is of order 1, 
 then in the CCKLP setup, the corrections to entanglement wedge reconstruction are exponentially small in $G$ relative to the
 gravitational backreaction -- measured by the dependence of the area term in the RT formula on the bulk quantum 
 state.   This nicely matches expectations in gravity.    Derivations of the RT formula  \cite{LM,Der1,Der2} are 
 based on semiclassical reasoning, which is expected to be valid to all orders of perturbation theory but not beyond, and 
 entanglement wedge reconstruction is expected to have the same degree of validity.\footnote{In particular, a simple and elegant argument showing that entanglement wedge reconstruction cannot be exact was given in \cite{Kelly}.
Consider a bulk operator $\phi$ that can be reconstructed by a boundary operator $\h\phi_A$ supported in a boundary region $A$, and can also be reconstructed by a boundary operator
$\h\phi_B$ supported in a boundary region $B$, but cannot be reconstructed in the intersection $A\cap B$; suppose further that $A\cup B$ is not the entire boundary.  Then $\h\phi_A\not=\h\phi_B$ (since if $\h\phi_A=\h\phi_B$, then this operator is actually
supported in $A\cap B$, contradicting the assertion that $\phi$ cannot be reconstructed in $A\cap B$), so $\h\phi_A-\h\phi_B\not=0$. If $\Omega$ is the ground state of the system and
 entanglement wedge reconstruction is exact, then $\phi\ket{\Omega}=\h\phi_A\ket{\Omega}=\h\phi_B\ket{\Omega}$, so 
$(\h\phi_A-\h\phi_B)\ket{\Omega}=0$, contradicting the Reeh-Schlieder theorem in the boundary CFT; this theorem  states that a nonzero operator supported in a proper subset of the boundary, such as $A\cup B$,
cannot annihilate the ground state.}    But gravitational backreaction on the geometry is expected
 to be perturbative.
 
As already noted,  in the CCKLP setup, there is no 
 area operator, only an area function on the set of quantum states. If the area $\sf A$ that appears in the RT formula is assumed to be the expectation of an area operator $\A$, then
  this is consequential for entanglement wedge reconstruction \cite{Harlow}.
 Suppose that the area of an RT surface that bounds a spatial region $a$ and also its spacelike complement $\bar a$ is a well-defined observable
  at the quantum 
 level.  If so, it can presumably be 
 measured both by an observer in $a$ and by an observer in $\bar a$.    Therefore, a hypothetical area operator $\A$ is 
 contained in the algebra $W_a$ of observables in region $a$ and also in the corresponding algebra $W_{\bar a}$ of region $\bar a$.  As these algebras commute with each other,
 $\A$ is central in both algebras,
  and in particular if
 $\A$ is not a $c$-number then these algebras  have nontrivial centers.   Once one assumes the existence of such  a nontrivial
 area operator $\A$ that is in the center of both algebras, there is no problem, in principle, with exact entanglement wedge reconstruction, 
 as detailed in \cite{Harlow}.     Should one expect the existence of such an area operator?  
 In perturbation theory in $G$,
 in continuum quantum field theory,  ultraviolet divergences  prevent the definition of an area operator $\A$ that is central in the
 two algebras.\footnote{In general, in quantum field theory in $D$ dimensions, it is not possible to define an operator by smearing a local field on
 a spacelike manifold $\Sigma$ of codimension 2.   A free scalar field $\phi$ has a short distance singularity $\phi(x)\phi(y)\sim 1/|x-y|^{D-2}$, which is not integrable in
 $D-2$ dimensions, so smearing it on  $\Sigma$  does not define an operator (if $\O_f=\int_\Sigma {\mathrm{d}}^{D-2}x\,f(x)\phi(x)$, for some smooth function $f$,
 and $\Omega$ is a normalizable state, then
$\O_f \Omega$ is unnormalizable, so $\O_f$ does not make sense as an operator).  
 The Lehmann-K\"{a}ll\'en spectral representation shows
 that any quantum field has a short distance singularity at least as severe as a free scalar, and hence serves no better.    So it is not possible, in quantum field theory in a fixed gravitational
 background,  to define
 an operator by smearing on a codimension 2 spacelike surface.  Perturbative gravity is similar.}   However, it is imaginable that a nonperturbative theory of quantum gravity would provide a cutoff
 that would obviate such issues.   And regardless,  in a cutoff approximation in which the state spaces  in  bulk regions $a$ and $\bar a$ are
finite-dimensional, it may be reasonable to assume the existence of such an operator.      At any rate, in the present article we work in the CCKLP framework with an area function but no area operator.\footnote{In the continuum limit, a cutoff area operator with some smearing in directions orthogonal to the RT surface presumably can be defined, at least in perturbation theory, but would not be central and so would not be a tool for exact entanglement wedge reconstruction.}
 
 Section 2 of this paper contains a review of relevant parts of the CCKLP construction, and a qualitative explanation of why one would expect in this framework that the
 errors in entanglement wedge reconstruction would be exponentially small compared to the backreaction on the geometry.   Section 3 contains a technical
 calculation\footnote{Anthropic's Claude Opus 4.8  helped in generalizing this calculation beyond some simple special cases that did not require assistance.}    demonstrating this statement.

 \section{Exponentially Small Errors in Entanglement Wedge Reconstruction}
 
 First we briefly recall the CCKLP setup.   In the boundary CFT, we consider a pair of complementary, spacelike separated regions $A$ and $\bar A$.    
 We assume that the CFT Hilbert space correspondingly factorizes as $\H=\H_A\otimes \H_\bA.$   We assume that  $\H$ has a ``code subspace''  
 isomorphic to the Hilbert space  $\H_L$ of the bulk low energy effective field theory \cite{ADH,HaPPY}.   We further assume $\H_L$ to factorize as 
 $\L_a\otimes \L_{\bar a}$, where $a$ and $\bar a$ are the entanglement wedges of $A$ and $\bA$,
 and $\L_a, $ $\L_{\bar a}$ are the corresponding Hilbert spaces. AdS/CFT duality is assumed to define isomorphisms mapping  $\L_{a}$ and $\L_{\bar a}$, respectively, to subspaces $\H_{A_1}$ and 
 $\H_{\bA_1}$ of $\H_A$ and $\H_{\bA}$; generally, we just identify the equivalent subspaces, $\L_a\cong \H_{A_1}$, $\L_{\bar a}\cong \H_{\bA_1}$.
 We assume that $\H_A$ and $\H_{\bA}$ have factorizations $\H_A=\H_{A_1}\otimes \H_{A_2}$, $\H_{\bA}=\H_{\bA_1}\otimes \H_{\bA_2}$, where in each case the first factor is a space of ``low energy'' states and the second describes ``short distance'' degrees of freedom responsible for the geometric entropy.
 
 AdS/CFT duality is assumed to involve an isometric ``encoding'' $V:\H_L\to \H$ of the low energy bulk Hilbert space $\H_L$ in the CFT Hilbert space $\H$.
 Consider a state $\psi\in \H_L$.  Restricted to an entanglement wedge $a$ or $\bar a$, $\psi$ can be represented by density matrices $\sigma_a=\Tr_{\bar a}\ket{\psi}\bra{\psi}$, $\sigma_{\bar a}=\Tr_a \ket{\psi} \bra{\psi}$.    Entanglement wedge reconstruction means that from the encoded state $V\psi$, an observer in region $A$
 can recover the low energy bulk density matrix $\sigma_a$ by applying a unitary transformation $U_A$ that acts only on $\H_A$; and similarly, an observer in region 
 $\bar A$ can recover $\sigma_{\bar a}$ by applying a unitary transformation $U_{\bA}$ that acts only on $\H_\bA$.
 
 Exact entanglement wedge reconstruction is very restrictive.   It is possible only if $V$ has the very special form
 \be\label{specform} V\psi=R_A{}^{-1} R_{\bar A}{}^{-1}(\psi\otimes\chi),\ee
 where $R_A$ and $R_{\bar A}$ are some unitary operators on $\H_A$ and $\H_{\bar A}$, $\chi$ is some fixed element of $\H_{A_2}\otimes \H_{\bA_2},$
 and we use the isomorphism $\H_L\cong \H_{A_1}\otimes \H_{\bar A_1}$ to view $\psi$ as an element of $\H_{A_1}\otimes \H_{\bar A_1}$.  
  If $V$ does have this form, an observer in region $A$ can recover $\sigma_a$ by applying the unitary transformation  $R_A$ to get $R_{\bar A}^{-1}\psi\otimes\chi$.  The density matrix of this state for region $A$ is $\sigma_a\otimes \Tr_{\bA_2}\ket{\chi}\bra{\chi}$, so  in this way the observer in region $A$ recovers the desired density matrix $\sigma_a$ of the entanglement wedge $\a$, tensored with a fixed density matrix $ \Tr_{\bA_2}\ket{\chi}\bra{\chi}$ of the high energy degrees of freedom.   Similarly, the observer in region $\bA$ can recover the density matrix $\sigma_{\bar a}$ of the entanglement wedge $\bar a$ by acting with $R_{\bA}$.
  The unitary operators $R_A$, $R_\bA$ that appear in this discussion are conveniently called ``local unitaries'' as they each act on one Hilbert space factor
  $\H_A$ or $\H_\bA$ associated to a particular region in spacetime.

 In this case of exact entanglement wedge reconstruction, we can easily obtain an RT-like formula for the entropy $S_A$ of the density matrix
 \be\label{zelbor}\sigma_A =\Tr_{\bar A} \ket{V\psi}\bra{V\psi}=\sigma_a\otimes   \Tr_{\bA_2}\ket{\chi}\bra{\chi}.\ee
   Write $S_A$ and $S_a$ for the entropies of the density matrices $\sigma_A $ and $\sigma_a$, and write $S_\chi$ for the entropy of the density matrix
 $ \Tr_{\bA_2}\ket{\chi}\bra{\chi}$ on $\H_{A_2}$.   Thus $S_\chi$, which is also the entropy of the density matrix  $ \Tr_{A_2}\ket{\chi}\bra{\chi}$ on $\H_{\bA_2}$,
 is the entanglement entropy between systems $A_2$ and $\bA_2$ in the state $\chi$.   The tensor product form of $\sigma_A$ in eqn. (\ref{zelbor}) implies that
 \be\label{elbo} S_A=S_a+S_\chi.  \ee   As noted in \cite{Harlow} in a related context, this result can be viewed as an analog of the RT formula,
 extended to include the bulk entropies as in \cite{Der1,Der2}.   That formula reads
 \be\label{yelbo}S_A=S_a+\frac{\sf A}{4G}. \ee
 So $S_\chi$, which is the entropy of some unspecified ``short distance'' bulk 
 degrees of freedom, plays the role of the area term in the usual formula.   But, since exact entanglement wedge 
 reconstruction is only possible if $\chi$ is a fixed state, independent of the state $\psi$ of the low energy degrees of freedom, 
 we see that in this framework, the area term in the entropy is a constant, 
 independent  of the quantum state of the low energy degrees of freedom.     
 
 If $V$ does not have the special form of eqn. (\ref{specform}), then an observer in region $A$ cannot precisely recover the bulk density matrix $\sigma_a$.  
 How well can one do?   As in \cite{CCKLP},  we will generalize the question slightly before discussing the optimal recovery.   Suppose that what one wishes
 to reconstruct is not a pure state $\psi\in \H_L$ but more generally a density matrix $\rL$ on $\H_L$.   The encoding maps this to $V\rL V^\dagger$, 
 and observers in regions $A$ and $\bar A$, by acting with unitary operators $R_A$, $R_{\bar A}$ on their parts of the state, and then tracing out the short distance degrees of
 freedom in $\H_{A_2}\otimes \H_{\bA_2}$,   can transform this to  
 \be\label{belbo}\sigma_{{\mathrm R}}=\Tr_{A_2\bA_2}R_A R_{\bA}V\rL V^\dagger R_A^\dagger R_\bA^\dagger.\ee
This operation defines a quantum channel $\sigma_{\mathrm R}=\N_{R_A,R_{\bar A}}(\rL)$.      CCKLP propose that  the ``optimal recovery'' is obtained
by choosing the local unitaries 
$R_A$, $R_{\bA}$ to maximize the coherent information\footnote{The coherent information of a quantum channel $\N$ mapping density matrices
on a Hilbert space $\H_L$ to density matrices on a possibly different Hilbert space $\H'$ is defined as follows.  Let $\H_r$ be a reference Hilbert space of the 
same dimension as $\H_L$, and define a maximally mixed state $\Phi$ on $\H_L\otimes \H_r$.   Consider the quantum channel $\N\otimes {\mathrm {Id}}_r$
mapping density matrices on $\H_L\otimes \H_r$ to density matrices on $\H'\otimes \H_r$.  Let $\sigma_{\H'\H_r}=(\N\otimes {\mathrm {Id}}_r)(\ket{\Phi}
\bra{\Phi})$, $\sigma_{\H'}=\Tr_{\H_r}\sigma_{\H'\H_r}$.   The coherent information of the channel is then $I_c(\N)=S(\sigma_{\H'})-S(\sigma_{\H'\otimes \H_r})$.}
of the channel $\N_{R_A,R_\bA}$.   The choice of this specific criterion for the optimal recovery will actually not play an important role in the present article,\footnote{This criterion optimizes the recovery of the full low energy state $\rL$ through the joint efforts of observers in the two regions $A$ and $\bar A$,
each acting separately on their own part of the state.   One would get a different answer if one chooses $R_A$ to optimize the recovery of $\sigma_a$, while  separately
choosing $R_{\bar A}$ to optimize the recovery of $\sigma_{\bar a}$.} for reasons
explained in section \ref{indep}.    We will abbreviate
the optimal pair $R_A, \,R_\bA$ as $R$,  denote the corresponding channel as $\N^{R}$,  and write $\sigma^{(R)}_{A_1 A_2}=\N^{R}(\rL)$
for the optimally recovered state on $A_1 A_2$. The optimally recovered state on $A_1$ or $A_2$ alone is then 
$\sigma^{(R)}_{A_1}=\Tr_{A_2} \sigma^{(R)}_{A_1 A_2}$.   We also define $\sigma^{(R)}_{A_2}=\Tr_{A_1} \sigma^{(R)}_{A_1 A_2}$ (this is simply
called $\chi$ in \cite{CCKLP}).   Optimal recovery on $\bar A=\bar A_1\bar A_2$ is defined similarly by tracing over $A$.

CCKLP defined the geometric entropy, which they also called the proto-area entropy $S_\PA$,
 to be the difference between  the entropy
$S_A=S(\sigma_A)$ of the full CFT state restricted to the boundary region $A$ and the entropy $S(\sigma^{(R)}_{A_1})$ of the reconstructed low energy state:\footnote{$S(\sigma_A)$ is unaffected by local unitaries acting separately on $A$ and $\bA$, so we write simply $S(\sigma_A)$ rather than
$S(\sigma_A^{(R)})$.}
\be\label{zilcox} S_\PA(A)=S(\sigma_A)-S(\sigma_{A_1}^{(R)}). \ee
In other words, the geometric entropy  is defined to be the difference between the entropy $S(\sigma_A)$ of the CFT state reduced to region $A$ and
the portion of this entropy that  a boundary observer  in region $A$ can interpret as the entropy of the reconstructed state $\sigma_{A_1}^{(R)}$.

As noted by CCKLP at the end of section 3 of \cite{CCKLP}, a drawback of this definition is the following.    Suppose that the bulk low energy fields are in a pure
state, $\rL=\ket{\psi}\bra{\psi}$ for some $\psi$.     Purity of the bulk state implies that the boundary CFT state is likewise pure, 
so the two boundary regions 
have the same entropy, $S(\sigma_A)=S(\sigma_{\bar A})$.  Purity of the bulk state also implies that  $S(\sigma_a)=S(\sigma_{\bar a})$, but since
there is no simple relation between the reconstruction errors in region $A$ and those in region $\bA$, generically $S(\sigma_{A_1}^{(R)})\not=
S(\sigma_{\bA_1}^{(R)})$, and therefore $S_\PA(A)\not= S_\PA(\bA)$.   This contradicts the fact that if the bulk state is pure,  one expects the two complementary boundary regions 
$A$ and $\bA$ to have the same RT surface, and hence the same geometric entropy. 

An alternative definition of the geometric entropy would be to subtract from the CFT entropy $S(\sigma_A)$ the actual bulk entropy $S(\sigma_a)$,
rather than the reconstructed bulk entropy $S(\sigma_{A_1}^{(R)})$.  Thus a modified version of the proto-area entropy would be
\be\label{nilcox} \t S_\PA(A)=S(\sigma_A)-S(\sigma_a). \ee
If the bulk state is pure, then $\t S_\PA(A)=\t S_\PA(\bA)$, since $S(\sigma_a)=S(\sigma_{\bar a})$ and $S(\sigma_A)=S(\sigma_\bA)$.

For definiteness we will continue with the definition of eqn. (\ref{zilcox}).   But we will see that in a certain sense the 
choice between these two definitions does not matter.
The difference between the two definitions is $S(\sigma_{A_1}^{(R)})-S(\sigma_a)$, which is a measure of the error in entanglement wedge reconstruction.
We will see that in  a situation that is natural in gravity, with $S(\sigma_a)\sim 1$ but $S(\sigma_A)\sim \frac{1}{G}$, the error in entanglement wedge 
reconstruction is exponentially small in $G$.    So up to exponentially small corrections, the two formulas for the geometric
entropy coincide.  

To understand the effects of small departures from exact entanglement wedge reconstruction, CCKLP considered the effects of replacing an 
encoding map $V$ that allows exact entanglement wedge reconstruction with $V_\epsilon=e^{\i \epsilon W} V$, where $\epsilon$ is a small real parameter and
$W$ is drawn from a GUE ensemble, more precisely, a Gaussian unitary ensemble of random hermitian matrices acting on $\H=\H_A\otimes \H_{\bA}$.   Thus rather than (\ref{specform}), the encoding map is now
\be\label{zencoding} V_\epsilon\psi = e^{\i\epsilon W} R_A^{-1} R_{\bar A}^{-1} (\psi\otimes \chi). \ee    Nothing essential changes if we replace $V_\epsilon$
by $R_A R_{\bar A}V_\epsilon = R_A R_\bA   e^{\i\epsilon W} R_A^{-1} R_{\bar A}^{-1} (\psi\otimes \chi), $  since the reconstruction is anyway supposed to
be optimized with respect to acting on the state with unitary operators on $\H_A$ and on $\H_\bA$, and the factors $R_A$, $R_\bA$ can be absorbed in that optimization.    And it also makes no difference to replace
$ R_A R_\bA   e^{\i\epsilon W} R_A^{-1} R_{\bar A}^{-1}$ by $e^{\i \epsilon W}$, since the ensemble from which $W$ is drawn is invariant under conjugation
by a unitary matrix.   So finally, to study small generic violations of entanglement wedge reconstuction, it suffices to consider the family of encoding maps
\be\label{wencoding} V_\epsilon \psi = e^{\i \epsilon W}\ket{\psi}\otimes \ket{\chi}.\ee

CCKLP primarily established two facts about such encodings.   First, in section 4, they show roughly that
the geometric entropy is a monotonically increasing function of
the bulk entropy.   (The precise relation of this statement to expectations in gravity appears to be subtle.)
Second, they show in section 5 that the violation of entanglement wedge reconstruction
 is controlled by what in quantum information theory is called the ``nonlocal magic'' of the encoded state.

 In these analyses, a useful tool was a formula (Theorem 4.1 in \cite{CCKLP}) expressing the proto-area entropy in terms of relative entropies.
 We recall that the relative entropy between two density matrices $\t\sigma$ and $\sigma$ on the same Hilbert space is defined as $D(\t\sigma||\sigma)=\Tr\,\t\sigma\log\t\sigma
 -\Tr\,\t\sigma\log\sigma = - S(\t\sigma)-\Tr\,\t\sigma\log\sigma$.   We will write $\sigma_{A_1A_2}^{(\epsilon)}$ for the density matrix on $A_1 A_2$ for the code (\ref{wencoding})
 (with some choice of $W$ and with the optimal choice of local unitaries $R$), and similarly we define $\sigma_{A_1}^{(\epsilon)}=\Tr_{A_2}\,\sigma_{A_1A_2}^{(\epsilon)}$, $\sigma_{A_2}^{(\epsilon)}=\Tr_{A_1} \,\sigma_{A_1 A_2}^{(\epsilon)}$.    At $\epsilon=0$, these reduce to the density matrices for the trivial
 code $V(\psi)=\psi\otimes\chi$.   Those density matrices satisfy $\sigma_{A_1A_2}^{(0)}=\sigma_{A_1}^{(0)}\otimes \sigma_{A_2}^{(0)}$, and hence $\log \sigma_{A_1A_2}^{(0)}-\log \sigma_{A_1}^{(0)}-\log \sigma_{A_2}^{(0)}=0$.    From this follows the identity
 \be\label{welcox} \Tr\, \sigma_{A_1A_2}^{(\epsilon)}\log\sigma^{(0)}_{A_1 A_2}- \Tr\, \sigma_{A_1}^{(\epsilon)}\log\sigma^{(0)}_{A_1}- \Tr\, \sigma_{A_2}^{(\epsilon)}\log\sigma^{(0)}_{A_2}=0.\ee
Since the von Neumann entropy of a density matrix is defined by $S(\rho)=-\Tr\,\rho\log\rho$,  the definition of the proto-area entropy is equivalent to
 \be\label{pelcox} S_\PA(A)= -\Tr\,\sigma^{(\epsilon)}_{A_1A_2}\log \sigma^{(\epsilon)}_{A_1A_2}+\Tr\,\sigma^{(\epsilon)}_{A_1}\log \sigma^{(\epsilon)}_{A_1}. \ee
Adding to this the identity (\ref{welcox}) and using the definition of relative entropy, we get
 \be\label{delcox}S_\PA(A) = -D(\sigma_{A_1A_2}^{(\epsilon)}||\sigma_{A_1A_2}^{(0)})+ D(\sigma_{A_1}^{(\epsilon)}||\sigma_{A_1}^{(0)})
  -\Tr\,\sigma_{A_2}^{(\epsilon)}\log \sigma_{A_2}^{(0)}.  \ee
  
  To get a qualitative understanding of this formula, we can assume that the state $\chi$ of the short distance degrees of freedom in $A_2 \bA_2$ is maximally
  mixed.   In this case, $\log \sigma_{A_2}^{(0)}$ is a multiple of the identity, so the last term in eqn. (\ref{delcox}) is a multiple of $\Tr\,\sigma_{A_2}^{(\epsilon)}$.
  Since every density matrix has trace 1, that term in eqn. (\ref{delcox}) is actually independent of $\epsilon$ and simply equals $-\Tr\,\sigma_{A_2}^{(0)}\log
  \sigma_{A_2}^{(0)}=S_\chi$.   So in this situation, the proto-area entropy is
  \be\label{elcox} S_\PA(A)=S_\chi - D(\sigma_{A_1A_2}^{(\epsilon)}||\sigma_{A_1A_2}^{(0)})+ D(\sigma_{A_1}^{(\epsilon)}||\sigma_{A_1}^{(0)}).\ee
 The backreaction -- that is, the dependence of $S_\PA(A)$ on the state $\psi$ of the low energy bulk degrees of freedom -- is contained in the
 difference of relative entropies $ - D(\sigma_{A_1A_2}^{(\epsilon)}||\sigma_{A_1A_2}^{(0)})+ D(\sigma_{A_1}^{(\epsilon)}||\sigma_{A_1}^{(0)})$.  
 
 The case that $\sigma_{A_2}^{(0)}$ is maximally mixed is a natural example in the CCKLP model.  In most of this article, we consider that case and use the formula (\ref{elcox}) for
 the geometric entropy.   However, in section \ref{general}, we relax the assumption that $\chi$ is maximally mixed and consider a general case.  At that point we will have to return to
 the more general formula of eqn. (\ref{delcox}).

 Now we can understand intuitively the assertion that in the CCKLP setup, the error in entanglement wedge reconstruction is exponentially small in $S_\chi$
 relative to gravitational backreaction.     Let $d_1$ and $d_2$ be the Hilbert space dimensions of $A_1$ and $A_2$.  To model gravity, we take $d_1$ to be large but independent of $G$,
while $d_2$ is exponentially large for small $G$:  $d_2\sim e^{S_\chi}\sim e^{c/G}$ (with $c>0$).   So in particular $\log d_1\ll \log d_2$.
 The backreaction involves the difference of two relative entropies $D(\sigma_{A_1A_2}^{(\epsilon)}||\sigma_{A_1A_2}^{(0)})$ and 
 $D(\sigma_{A_1}^{(\epsilon)}||\sigma_{A_1}^{(0)})$.     The first of these measures the distinguishability of two density matrices 
 $\sigma_{A_1 A_2}^{(\epsilon)}$ and $\sigma_{A_1A_2}^{(0)}$.
 The second measures the distinguishability of the same two states after restricting to $A_1$  by taking a partial trace over $A_2$.   Monotonicity of relative entropy
 under partial trace says that one always has $D(\sigma_{A_1}^{(\epsilon)}||\sigma_{A_1}^{(0)})<D(\sigma_{A_1A_2}^{(\epsilon)}||\sigma_{A_1A_2}^{(0)})$, but
 more specifically in the present case, for $\log d_1\ll \log d_2$, one expects 
 \be\label{belcox} D(\sigma_{A_1}^{(\epsilon)}||\sigma_{A_1}^{(0)})\ll D(\sigma_{A_1A_2}^{(\epsilon)}||\sigma_{A_1A_2}^{(0)}).\ee
On the right hand side, one is comparing two density matrices, and on the left hand side one is comparing the same two density matrices
after restricting to an exponentially small part of the state.    The restriction loses almost all of the information about the difference between
the two density matrices, so  one expects the left hand side of eqn. (\ref{belcox}) to be exponentially 
smaller than  the right hand side.     Hence, to within very small errors, the backreaction of the bulk state on the geometric entropy is 
given by just one term on the right hand side of eqn. (\ref{elcox}), namely $- D(\sigma_{A_1A_2}^{(\epsilon)}||\sigma_{A_1A_2}^{(0)})$.    The other
term  is negligible.

On the other hand, the left hand side of eqn. (\ref{belcox}) measures the error in entanglement wedge reconstruction.   Indeed, $\sigma_{A_1}^{(0)}$ is the
low energy bulk state $\sigma_a=\Tr_{\bar a}\,\ket{\psi}\bra{\psi}$, and $\sigma_{A_1}^{(\epsilon)}$ is the reconstructed version of this state for the encoding map (\ref{wencoding}).
The relative entropy between two states is a useful measure of the difference between them.
So the inequality (\ref{belcox}) is the statement that the error in entanglement wedge reconstruction is vastly smaller than the backreaction of the bulk quantum
state on the geometry.

The two definitions of the geometric entropy in eqn. (\ref{zilcox}) and eqn. (\ref{nilcox}) differ by $S(\sigma_{A_1}^{(\epsilon)})-S(\sigma_{A_1}^{(0)})$.
Standard inequalities in quantum information theory bound this in terms of the relative entropy $ D(\sigma_{A_1}^{(\epsilon)}||\sigma_{A_1}^{(0)})$ between
the two density matrices, and therefore if it is true that that relative entropy is nonperturbatively small, then the same is true for the difference between the
two possibilities for the geometric entropy.

We have implicitly assumed that nothing is special about the small subspace $\H_{A_1}$ to which the density matrices are restricted
before defining the relative entropy between them that appears on the left hand side of eqn. (\ref{belcox}).   Indeed, in the CCKLP setup, the matrix $W$ in the definition
(\ref{wencoding}) of the encoding map is assumed to be a Gaussian random matrix on the full Hilbert space $\H=\H_A\otimes \H_\bA$, with no role for
the decompositions $A=A_1A_2$, $\bA=\bA_1\bA_2$.

A technical calculation justifying the preceding  heuristic claims and filling in some gaps is the subject of  the next section.  Among other things, 
we will explain why the optimization over local unitaries $R_A$, $R_\bA$, which we have ignored in this discussion, actually does not play an important role.

\section{Technical Calculation}

\subsection{Preliminaries}\label{preliminaries}

The goal of this section is to present a technical calculation to make quantitative the inequality (\ref{belcox})
which asserts the smallness of the errors in entanglement wedge reconstruction.  We work throughout to lowest nontrivial order in the perturbation
parameter $\epsilon$ of eqn. (\ref{wencoding}), namely second order.    As before, we set $d_1={\mathrm{dim}}\,\H_{A_1},$ $d_2={\mathrm{dim}}\,
\H_{A_2}$.   
For brevity, we define
 \be\label{turbex}D_{A_1}= D(\sigma_{A_1}^{(\epsilon)}|\sigma_{A_1}^{(0)}),~~~~~D_A=D_{A_1A_2}=D(\sigma_{A_1A_2}^{(\epsilon)}||\sigma_{A_1A_2}^{(0)}).\ee
To begin with, we will assume that the short distance states are described by a maximally mixed pure state  $\chi\in\H_{A_2}\otimes \H_{\bar A_2}$,
with entropy $S_\chi=\log d_2$.  Under this assumption, we will show that the errors in entanglement wedge reconstruction are of order $\frac{1}{d_2^2}$ relative
to the gravitational backreaction.      But actually, we will learn in section \ref{general} that relaxing this assumption does not change the qualitative result.

  To begin with, we will assume that the  low energy bulk fields are also in a pure state $\psi\in \H_{A_1}\otimes
\H_{\bar A_1}$.   After analyzing that case, the generalization to the case that the  low energy bulk 
state is mixed will be described more briefly in section \ref{mixed}.   Also, to begin with we will ignore the fact that the reconstruction is supposed to be optimized over the choice of local unitaries $R_A$, $R_{\bar A}$.   We will explain in section \ref{indep}  why this optimization does not affect the main conclusion.

For applications to gravity, one expects that in the state $\psi$, there is a large entanglement entropy between regions $A_1$ and $\bA_1$, because of entanglement of the low energy fields across the RT surface.   In fact, the Reeh-Schlieder theorem of quantum field theory (and its extension to curved spacetime \cite{StrohWitten})
implies that in a nice class of states, including those that have a simple Euclidean preparation, the density matrices $\sigma_{A_1}^{(0)}$, $\sigma_{\bA_1}^{(0)}$ are invertible.   
We will assume this and in particular assume that $\H_{A_1}$ and $\H_{\bA_1}$ have the same dimension $d_1$.
Actually, because of the logarithms in the definition of the relative entropy,
the relative entropy cannot be expanded in powers of $\epsilon$ unless  $\sigma_{A_1}^{(0)}$ is  invertible, and the coefficients in the expansion are
anomalously large if that density matrix is invertible but has exponentially small eigenvalues (see the discussion of eqn. (\ref{firstgoal})).  Exponentially small eigenvalues are realistic in quantum field theory, because of massive particles that can be localized far from the entangling surface.   However, if one wants to take exponentially small eigenvalues into account, then one probably should refine the model of \cite{CCKLP}, which is based on a Gaussian unitary ensemble in which all modes are treated equally, to suppress  contributions of the modes with exponentially small eigenvalues.    In the context of the model of \cite{CCKLP} which we study in the present article, it is reasonable to assume that the eigenvalues of the density matrix of the low energy fields are not exponentially small.

\subsection{Second-Order Relative Entropy}\label{secondorder}

Consider in general a density matrix $\sigma$ with eigenvectors $\ket{u}$ and eigenvalues $r_u$, and a small perturbation to another density matrix $\sigma+\delta$;
here $\delta$ is a hermitian matrix with trace zero.   The relative entropy $D(\sigma+\delta||\sigma)$ vanishes at $\delta=0$ and also  at first
order in $\delta$.   In second order in $\delta$, setting $\delta_{uu'}=\bra{u'}\delta\ket{u}$, the relative entropy is given by a formula of Kubo and Mori:
\begin{equation}\label{KM}
D\big(\sigma+\delta\,\big\|\,\sigma\big)
=\tfrac12\sum_{u,u'}w(r_u,r_{u'})\,|\delta_{uu'}|^2+O(\delta^3),
\qquad w(r,r')=\frac{\log r-\log r'}{r-r'},
\end{equation}
where one also defines  $w(r,r)=1/r$.  Separating out the diagonal contributions, an equivalent formula for the leading term is
\begin{equation}
D=\tfrac12\sum_u\frac{|\delta_{uu}|^2}{r_u}+\sum_{u<u'}w(r_u,r_{u'})\,|\delta_{uu'}|^2 .
\end{equation}

\subsection{The Perturbation}

Because in general $D(\sigma+\delta||\sigma)$ is of order $\delta^2$ for small $\delta$, in order to analyze the inequality (\ref{belcox}) to order $\epsilon^2$, we only
need to know the perturbation to the density matrix to order $\epsilon$.   This can be worked out as follows.

The initially assumed pure state
 $\Phi=\psi\otimes \chi\in \H_A\otimes \H_\bA$ has a 
density matrix\footnote{We reserve the letter $\rho$ for a density matrix of the full system $A\bA$ and $\sigma$ for density matrices of subsystems.   For density matrices of
subsystems,
we indicate the subsystem by a footnote and whether $\epsilon$ is set to zero by a superscript, so we write $\sigma_{A_1}^{(0)}$, $\sigma_{A_1}^{(\epsilon)}$, etc.
For the full system, we write simply $\rho$ for the density matrix at $\epsilon=0$ and (on the few occasions where it appears) $\rho_\epsilon$ for the extension to $\epsilon\not=0$.}
$\rho=\ket{\Phi}\bra{\Phi}$ on $\H_A$.  The perturbed encoding map (\ref{wencoding}) replaces $\Phi$ with   
$\Phi_\epsilon=e^{\i \epsilon W}\Phi$. The corresponding density matrix $\rho_\epsilon=\ket{\Phi_\epsilon}\bra{\Phi_\epsilon}$ is 
\be\label{pertden} \rho_\epsilon=e^{\i\epsilon W}\rho e^{-\i \epsilon W}=\rho+\i\epsilon[W,\rho]+\O(\epsilon^2).\ee
 We set $M=[W,\rho]$.

If $S$ is any subsystem of $A\bar A$ with complement $S^c$, then the reduced state on $S$ is $\sigma_S=\Tr_{S^c}\,\rho$, and its first order variation is
\be\label{erden}\delta \sigma_S=\i \epsilon\,\Tr_{S^c}\, M \ee
We will need two cases of this, with $S=A=A_1 A_2$, $S^c=\bar A$, or $S=A_1$, $S^c=A_2\bar A$.

\subsection{The GUE Second Moment}

A Gaussian random matrix $W$ with distribution proportional to $\exp(-\Tr\,W^2/2\sigma_W^2)$ has second moment 
\begin{equation}\label{WW}
\mathbb \E[W_{IJ}W_{KL}]=\sw^2\,\delta_{IL}\delta_{JK}.
\end{equation} ($\E[\,\cdot\,]$ will denote an expectation value with respect to the Gaussian measure, and 
capital indices
 $I,J,K,L$ will range over an orthonormal basis of $\H=\H_A\otimes \H_\bA$.   Since states and density matrices we encounter can be assumed real,
we do not distinguish covariant and contravariant indices.)  We want to convert (\ref{WW}) to a formula for $\E[M_{IJ}M_{KL}]$.  With $M_{IJ}=\sum_P(W_{IP}\rho_{PJ}-\rho_{IP}W_{PJ})$, we get
\begin{equation}
M_{IJ}M_{KL}=\sum_{P,Q}\Big[
\underbrace{W_{IP}\rho_{PJ}W_{KQ}\rho_{QL}}_{1}
-\underbrace{W_{IP}\rho_{PJ}\rho_{KQ}W_{QL}}_{2}
-\underbrace{\rho_{IP}W_{PJ}W_{KQ}\rho_{QL}}_{3}
+\underbrace{\rho_{IP}W_{PJ}\rho_{KQ}W_{QL}}_{4}\Big].
\end{equation}
Applying \eqref{WW} to the two $W$'s in each term, we get
\begin{align}\notag
1:&\quad \mathbb E[W_{IP}W_{KQ}]\rho_{PJ}\rho_{QL}
=\sw^2\delta_{IQ}\delta_{PK}\rho_{PJ}\rho_{QL}=\sw^2\rho_{KJ}\rho_{IL},\\ \notag
2:&\quad -\mathbb E[W_{IP}W_{QL}]\rho_{PJ}\rho_{KQ}
=-\sw^2\delta_{IL}\delta_{PQ}\rho_{PJ}\rho_{KQ}=-\sw^2\delta_{IL}(\rho^2)_{KJ},\\ \notag
3:&\quad -\mathbb E[W_{PJ}W_{KQ}]\rho_{IP}\rho_{QL}
=-\sw^2\delta_{JK}\delta_{PQ}\rho_{IP}\rho_{QL}=-\sw^2\delta_{JK}(\rho^2)_{IL} \\
4:&\quad \mathbb E[W_{PJ}W_{QL}]\rho_{IP}\rho_{KQ}
=\sw^2\delta_{PL}\delta_{JQ}\rho_{IP}\rho_{KQ}=\sw^2\rho_{IL}\rho_{KJ} 
\end{align}
and therefore
\begin{equation}\label{MM}
\boxed{\;\mathbb E[M_{IJ}M_{KL}]
=\sw^2\big(2\rho_{KJ}\rho_{IL}-\delta_{IL}(\rho^2)_{KJ}-\delta_{JK}(\rho^2)_{IL}\big).\;}
\end{equation}
If the bulk state of the low energy fields is pure, we can further set $\rho^2=\rho$, but the result (\ref{MM}) is valid without this
assumption.
 
 \subsection{From $\rho$ to $\sigma$: the Resolved Second Moment}
 
With  $\rho_\epsilon=\rho+\i\epsilon M$, eqn. (\ref{MM}) describes the second moment of the fluctuation in $\rho_\epsilon$ around its value at $\epsilon=0$.
What we actually need is not the second moment of  $\rho$, which  is a density matrix on the full Hilbert space $\H_A\otimes \H_{\bA}$, but the analogous second moment of a reduced density matrix on a subsystem $S$ of $A\bA$, with complement $S^c$.   More precisely, we need the second moment of $\delta \sigma_S$, defined in eqn. (\ref{erden}).    As remarked
earlier, $S$ will be either $A$ or $A_1$.

Write an $S$-index as $s$ and an $S^c$-index as $\bar s$. Using \eqref{erden} and hermiticity
$|(\delta\sigma_S)_{ss'}|^2=(\delta\sigma_S)_{ss'}(\delta\sigma_S)_{s's}$, we have
\begin{equation}\label{usefulone}
\mathbb E\bigl[|(\delta\sigma_S)_{ss'}|^2\bigr]
=-\eps^2\sum_{\bar s,\bar t}
\mathbb E\big[M_{(s\bar s)(s'\bar s)}\,M_{(s'\bar t)(s\bar t)}\big].
\end{equation}
To get this, one sets $I=(s\bar s),\,J=(s'\bar s),\,K=(s'\bar t),\,L=(s\bar t)$ in \eqref{MM} and carries out the desired partial traces by summing over  the
two $S^c$  indices $\bar s,\bar t$. 

 The $2\rho_{KJ}\rho_{IL}$ piece, for $\rho=\ket{\Phi}\bra{\Phi}$ pure, leads to
\begin{equation}
\sum_{\bar s,\bar t}\rho_{(s'\bar t)(s'\bar s)}\rho_{(s\bar s)(s\bar t)}
=\sum_{\bar s,\bar t}\Phi_{s'\bar t}\Phi^*_{s'\bar s}\,\Phi_{s\bar s}\Phi^*_{s\bar t}.
\end{equation}
The index $\bar s$ appears only in $\Phi^*_{s'\bar s}\Phi_{s\bar s}$ and $\bar t$ only in
$\Phi_{s'\bar t}\Phi^*_{s\bar t}$, so the double sum factorizes and gives
\begin{equation}
\Big(\sum_{\bar s}\Phi_{s\bar s}\Phi^*_{s'\bar s}\Big)
 \Big(\sum_{\bar t}\Phi_{s'\bar t}\Phi^*_{s\bar t}\Big)
=(\sigma^{(0)}_S)_{ss'}\,(\sigma^{(0)}_S)_{s's}
=\big|(\sigma^{(0)}_S)_{ss'}\big|^2,
\end{equation}
using $(\sigma^{(0)}_S)_{ss'}=\sum_{\bar s}\Phi_{s\bar s}\Phi^*_{s'\bar s}$.

In evaluating the terms in \eqref{MM} proportional to $\delta_{IL}$ or $\delta_{JK}$, we will, for now, assume that the overall
state is pure and replace $\rho^2$ with $\rho$.  With the same substitutions of indices as before, we have
 $\delta_{IL}=\delta_{(s\bar s)(s\bar t)}=\delta_{\bar s\bar t}$.  This Kronecker delta restricts the sum over $\bar s,\bar t$ to
 $\bar t=\bar s$, after which we are left with $
\sum_{\bar s}\rho_{(s'\bar s)(s'\bar s)}=(\sigma^{(0)}_S)_{s's'}$; likewise $\delta_{JK}\rho_{IL}\to
(\sigma^{(0)}_S)_{ss}$.

Collecting the three contributions,
\begin{equation}\label{resolved}
\boxed{\;\mathbb E\bigl[|(\delta\sigma_S)_{ss'}|^2\bigr]
=\eps^2\sw^2\Big[(\sigma^{(0)}_S)_{ss}+(\sigma^{(0)}_S)_{s's'}
-2\big|(\sigma^{(0)}_S)_{ss'}\big|^2\Big].\;}
\end{equation}
This is basis independent and holds  for either $S=A$ or $S=A_1$. Summing over $s,s'$
gives $\mathbb E\bigl[\Tr_S\delta\sigma_S^2\bigr]=2\eps^2\sw^2\big(d_S-\Tr\,\sigma_S^2\big)$, where $d_S$ is the dimension of $S$.

 \subsection{Scaling of Weights}

Suppose that $\sigma_{A_1}^{(0)}$ has eigenstates $\ket{i}$ with eigenvalues $p_i$.
If $\chi$ is maximally mixed, then 
\begin{equation}
\sigma^{(0)}_{A_1A_2}=\sigma_{A_1}^{(0)}\otimes\frac{\mathrm 1}{d_2},
\end{equation}
and has eigenvalues  $p_i/d_2$ in a basis of the form  $\ket{i\mu}$, with $\ket{\mu}$ running over a basis of $\H_{A_2}$.

Dividing every eigenvalue by
$d_2$ scales  by $d_2$ the weights that appear in the Kubo-Mori formula:
\begin{equation}\label{rescale}
w\!\left(\frac{p_i}{d_2},\frac{p_j}{d_2}\right)
=\frac{\log p_i/d-\log p_j/d}{(p_i-p_j)/d_2}=d_2\,w(p_i,p_j).
\end{equation}
This will lead to a simple final answer, as we see momentarily.

\subsection{Combining the Pieces}\label{combining}

$\sigma_{A_1}^{(0)}$ is diagonal in the basis $\ket{i}$:   $(\sigma_{A_1}^{(0)})_{ij}   = p_i\delta_{ij}$.   Similarly $\sigma_A^{(0)}$ is diagonal in the
basis $\ket{i\mu}$: $(\sigma_A^{(0)})_{i\mu,j\nu}=\frac{p_i}{d_2}\delta_{ij}\delta_{\mu\nu}$.

So we can explicitly evaluate (\ref{resolved}):
\begin{align}
\mathbb E\left[|(\delta\sigma_{A_1})_{ij}|^2\right]
&=\eps^2\sw^2\big[(p_i+p_j)-2p_i^2\,\delta_{ij}\big].
\label{mA1}\\[2pt]
\mathbb E\left[|(\delta\sigma_{A_1A_2})_{(i\mu)(j\nu)}|^2\right]
&=\eps^2\sw^2\Big[\frac{p_i+p_j}{d_2}-\frac{2p_i^2}{d_2^2}\,\delta_{ij}\delta_{\mu\nu}\Big],
\label{mA}
\end{align}

 From \eqref{KM} and \eqref{mA1}, we calculate
\begin{equation}\label{firstgoal}
\mathbb E\left[D_{A_1}\right]
=\tfrac12\sum_{ij}w(p_i,p_j)\,\eps^2\sw^2\big[(p_i+p_j)-2p_i^2\delta_{ij}\big]
=\eps^2\sw^2\,(\Xi-1),
\end{equation}
where\begin{equation}\label{zolf}
\Xi=\tfrac12\sum_{ij}(p_i+p_j)\,w(p_i,p_j).
\end{equation}
To estimate the order of magnitude of $\Xi$, note that $\Xi=d_1^2$ if $\sigma_{A_1}^{(0)}$ is maximally mixed, since $w(p,p)=1/p$, and all contributions to the sum
in eqn. (\ref{zolf}) are equal to 1.    In fact,\footnote{\label{conv} One has $t\coth t\geq1$ for all real $t$ with equality only at $t=0$ 
(this is equivalent to $\tanh t<t$ for positive $t$).  So setting $t=x/2$, 
we learn that $L(x)=x\coth\frac{x}{2}$ obeys $L(x)\geq 2$ for all real $x$.   As $(r+r')w(r,r')=L(\log r-\log r')$, this yields the inequality $(r+r')w(r,r')\geq 2$.
Moreover $L$ is a convex function, since $L''(x)= \frac{1}{2}\csch^2\frac{x}{2}\left(x\coth\frac{x}{2}-2\right) =  \frac{1}{2}\csch^2\frac{x}{2}(L(x)-2)$, which is strictly positive for all
real $x$.  This convexity will be important in section \ref{general}.}
 $w(p,p')\geq \frac{2}{p+p'}$ for all $p,p'$, so in general all contributions to $\Xi$ are at least 1 and $\Xi\geq d_1^2$.
There is no upper bound on $\Xi$, since it is infinite if one of the $p_i$ vanishes and  extremely large if one of the $p_i$ is exponentially small.
As explained in section \ref{preliminaries}, within the range of validity of the CCKLP model, one probably should assume that $\sigma_{A_1}^{(0)}$ does not
have exponentially small eigenvalues. So typically $\Xi\sim d_1^2$.    But actually, for our aim of proving that $D_{A_1}\ll D_{A_1A_2}$, the value of $\Xi$ will not matter, since we will see that $D_{A_1A_2}$
is also proportional to $\Xi$.

Now we consider $D_{A_1A_2}$, using  \eqref{KM} and the rescaling \eqref{rescale}.
For large $d_2$, the leading contribution to $D_{A_1A_2}$ comes from the $\frac{p_i+p_j}{d_2}$ term in \eqref{mA} and is
\begin{equation}\label{zefto}
\tfrac12\sum_{i\mu,j\nu}\underbrace{d_2\,w(p_i,p_j)}_{\text{weight}}
\cdot\underbrace{\eps^2\sw^2\,\frac{p_i+p_j}{d_2}}_{\text{moment}}
=\eps^2\sw^2\;d_2^2\;\underbrace{\tfrac12\sum_{ij}w(p_i,p_j)(p_i+p_j)}_{\Xi}.
\end{equation}
After canceling factors of $d_2$ and $\frac{1}{d_2}$ that are explicitly written on the left hand side of  eqn. (\ref{zefto}), a factor of $d_2^2$ has come
from the sum over $\mu,\nu$.   
 The subleading contribution to $D_{A_1A_2}$ from the $-\frac{2p_i^2}{d_2^2}$ term in \eqref{mA} can also be evaluated nicely:
\begin{equation}
-\tfrac12\,\eps^2\sw^2\,d_2\cdot\frac{2}{d_2^2}\sum_i\frac1{p_i}p_i^2\sum_{\mu\nu}\delta_{\mu\nu}
=-\eps^2\sw^2\,\frac{d_2\cdot d_2}{d_2^2}\sum_ip_i=-\eps^2\sw^2 .
\end{equation}
Hence
\begin{equation}
\mathbb E[D_{A_1A_2}]=\eps^2\sw^2\,(d_2^2\,\Xi-1).
\end{equation}

In sum,
\begin{equation}\label{ratio}
\boxed{\;
\mathbb E[D_{A_1}]=\eps^2\sw^2(\Xi-1),\quad
\mathbb E[D_{A_1A_2}]=\eps^2\sw^2(d_2^2\,\Xi-1),\quad
\frac{\mathbb E[D_{A_1}]}{\mathbb E[D_{A_1A_2}]}=\frac{\Xi-1}{d_2^2\Xi-1}
\approx\frac{1}{d_2^2}\Big(1-\frac1\Xi\Big).
\;}
\end{equation}
The bulk state enters these formulas only through the function  $\Xi$.   Since $\Xi$ is always large, the ratio $\E[D_{A_1}]/E[D_{A_1A_2}]$ is always close to
$\frac{1}{d_2^2}$, showing that the errors in entanglement wedge reconstruction are very small compared to the gravitational backreaction.   
 
We have expressed these results in terms of average values $\E[D_{A_1}]$ and $\E[D_{A_1A_2}]$, but actually $D_{A_1}$ and $D_{A_1A_2}$ are self-averaging,
meaning that they  take almost the same value for almost every draw of $W$ from the Gaussian ensemble.   For $X=A_1$ or $A_1A_2$, this  can be proved by calculating the standard deviation 
$\Delta_X$, defined by
$\Delta_X^2= \E[D_X^2]-\E[D_X]^2$.     For $d_2\gg d_1\gg 1$, the standard deviations are small, $\Delta_X\ll \E[D_X]$, for each choice of $X$.  In fact, $\frac{\Delta_{A_1A_2}}{\E[D_{A_1A_2}]}\sim \frac{1}{d_1d_2}$, $\frac{\Delta_{A_1}}{\E[D_{A_1}]}\sim \frac{1}{d_1}$.   These results can be found by expanding in powers of $\epsilon W$ up to fourth order.

 \subsection{General  ``High Energy''  Density Matrix}\label{general}
 
Here we will relax the assumption that the state $\chi$ on $A_2 \bA_2$ is maximally mixed. 
That means that we have to use the general formula (\ref{delcox}) for the relative entropies; this formula contains an extra term 
$Y=-\Tr\,\sigma_{A_2}^{(\epsilon)}\log\sigma_{A_2}^{(0)}$ that we have not had to consider so far.   Before analyzing this term, 
we consider how the relative entropies depend on $\sigma_{A_2}^{(0)}$, 
if we do not assume it to be maximally
mixed.

For general $\chi$, we can write
 \be\label{belfo}\sigma_{A_2}^{(0)}= \frac{m}{d_2}.\ee  In general,  $m$ is an arbitrary 
 positive matrix with eigenvalues $m_\mu$ and $\Tr\,m=d_2$. But in the context of the CCKLP model, one would not expect extremely large or extremely small eigenvalues of
 $m$ to be important.  One would not want $m$ to have extremely small eigenvalues, because the modes contributing to the geometric entropy are presumed to be highly entangled,
 and one does not want a major contribution to $\Tr\,m=d_2$ to come from exceptional large eigenvalues, as the geometric entropy is presumed to be a collective effect.
 
Dropping the assumption that $\sigma_{A_2}^{(0)}$ is maximally mixed actually has no effect on the derivation of
 $\E[D_{A_1}]$, because eqn. (\ref{mA1}) only involves the eigenvalues of $\sigma_{A_1}^{(0)}$, with $A_2$ traced out.  We do have to reconsider the derivation of
 (\ref{mA}), however.   Rather than $p_i/d_2$, the eigenvalues of $\sigma_{A_1A_2}^{(0)}$ are now $p_i m_\mu/d_2$.   As a result, in the derivation of eqn. (\ref{zefto}),
 the sum $\Xi$ is replaced by
 \be\label{zeftax}\Xi_A = \frac{1}{2 d_2^2}\sum_{(i\mu)(j\nu)} L(\log p_i-\log p_j+\log m_\mu -\log m_\nu), \ee
 where as in footnote \ref{conv}, $L(x)=x\coth\frac{x}{2}$, and we used $(r+r')w(r,r')=L(\log r-\log r')$.    Since $\sum_{\mu \nu} (\log m_\mu -\log m_\nu)=0$, Jensen's inequality, applied to the convex function $L$,
 implies that $\Xi_A\geq \Xi$, with equality precisely when $m=1$ and all $\log m_\mu-\log m_\nu$ vanish.  The upshot is that
 \be\label{eflax} \frac{\E[D_{A_1}]}{\E[D_{A_1A_2}]} =\frac{\Xi-1}{d_2^2 \Xi_A-1},~~~ \Xi_A\geq \Xi.\ee
 Thus the hierarchy of relative entropies does not depend on the assumption that $\chi$ is maximally mixed.

 Now we consider the extra term in the geometric entropy  that is present when $\chi$ is not maximally mixed.  This term has novel properties for the following reason.
 The  familiar relative entropies $D_{A_1}$ and $D_{A_1A_2}$ have the property that, even for a fixed draw $W$ from the Gaussian ensemble, if they are expanded in powers of
 $\epsilon$ around $\epsilon=0$, the first nonconstant term is quadratic.  The reason for this is that the relative entropy between nearby density matrices
 $\sigma$ and $\sigma+\delta$ is of order $\delta^2$ (a fact exploited in section \ref{secondorder}).    In contrast,  $Y$ does have a term of order $\epsilon$.   The  expansion of $Y$ begins
 \be\label{wongo}Y=S_\chi -\i\epsilon \Tr_{A_2}\left[\left(\Tr_{A_1\bA}[W,\rho]\right)\log\sigma_{A_2}^{(0)}\right]+
 \frac{\epsilon^2}{2}\Tr_{A_2}\left[\left(\Tr_{A_1\bA}[W,[W,\rho]]\right)\log\sigma_{A_2}^{(0)}\right]+\O(\epsilon^3),\ee where $\rho=\ket{\Phi}\bra{\Phi}$ is the density matrix of the full
 system $A\bA$, and omitted
  terms are negligible.   
  
  Just to  compute the expectation value $\E[Y]$, we can ignore the term of order $\epsilon$, since it is linear in $W$ and its ensemble
 average vanishes. Using (\ref{WW}), we can calculate $\E\bigl[[W,[W,\rho]]\bigr]$, with a very simple answer:
 \be\label{tilmo}  \E\bigl[[W,[W,\rho]]\bigr]=2\sigma_W^2 \left({\mathcal D} \rho -1\right),\ee
 where ${\mathcal D}=d_1^2 d_2^2$ is the dimension of the Hilbert space.
 This is traceless as expected, and leads to 
 \be\label{ilmo}\E\left[\Tr_{A_1\bA} [W,[W,\rho]]\right]= 2\sigma_W^2 d_1^2d_2(m-1)  \ee with $m$ defined as before.
 So
 \be\label{zongo}\E[Y]= \epsilon^2 d_1^2 d_2\sigma_W^2 \Tr_{A_2}(m-1)\log m. \ee
If eigenvalues of $m$ are not exceptionally large or small and $m$ is not too close to 1, then  $\E[Y]$  has a similar magnitude to  $D_{A_1A_2}$.  But importantly it
 is completely independent of the state $\psi$ of the low energy fields, so it does not contribute at all to the backreaction of the low energy fields on the geometry.
 It is a correction to the geometric entropy that is not sensitive at all to the low energy fields.
 
For a fuller picture, however, since $Y$ does have a term linear in $\epsilon$, we should compute the variance of $Y$.  For this, define $P=1_{A_1\bA}\otimes \log\sigma_{A_2}^{(0)}$.
Then the linear term in $Y$ is $Y_{(1)}= -\i\epsilon\Tr \left(W[\rho,P]\right)$.   So its variance is
\be\label{zelbo}\E[Y_{(1)} ^2]=-\epsilon^2 \E\left[ \left( \Tr \left(W[\rho,P]\right)\right)^2\right]=-\epsilon^2\sigma_W^2 \Tr\,[\rho,P]^2.\ee
So far in this article, we have assumed $\rho=\ket{\psi}\bra{\psi}\otimes \ket{\chi}\bra{\chi}$, where $\psi$ and $\chi$ are pure states respectively of the low energy fields and the
high energy degrees of freedom responsible for the geometric entropy.   However, with a view toward section \ref{mixed}, we will relax the assumption that the low energy fields
are in a pure state and replace $\ket{\psi}\bra{\psi}$ with a general density matrix $\rL$ of the low energy fields.   In the spirit of the CCKLP model, one assumes that the
high energy modes are frozen in some sort of  pure ``ground state'' $\chi$.   
So we take $\rho=\rL\otimes \ket{\chi}\bra{\chi}$.   Since $P$ acts nontrivially only
on $A_2$,  we have $\Tr_{A\bA}\,[\rho,P]^2=\Tr_{A_1\bA_1}\rL^2 \Tr_{A_2\bA_2}\left( \left[\ket{\chi}\bra{\chi},\log\sigma_{A_2}^{(0)}\right]^2\right)=2\Tr\,\rL^2 \left((\Tr\,\sigma_{A_2}\log\sigma_{A_2} )^2
-(\Tr\,\sigma_{A_2}\log^2\sigma_{A_2})\right)$.
  So
  \be\label{believeme} \E[Y_{(1)}^2] =2\epsilon^2\sigma_W^2\left( \Tr\,\rL^2\right) \left(  \sum_\mu \frac{m_\mu}{d_2}\log^2\frac{m_\mu}{d_2} -\left(\sum_\mu\frac{m_\mu}{d_2}\log\frac{m_\mu}{d_2}\right)^2\right).    \ee   The last factor is a measure of the dispersion in the eigenvalues of $\sigma_{A_2}^{(0)}$ and for our purposes is typically of order 1.
  The typical magnitude of $Y_{(1)}$ is the square root of this:
  \be\label{lieveme} \sqrt{\E[Y_{(1)}^2] }=2^{1/2}\epsilon\sigma_W  \sqrt{\Tr\,\rL^2} \left(  \sum_\mu \frac{m_\mu}{d_2}\log^2\frac{m_\mu}{d_2} -\left(\sum_\mu\frac{m_\mu}{d_2}\log\frac{m_\mu}{d_2}\right)^2\right)^{1/2}.\ee
  If the state $\psi$ of the low energy fields is pure, then $\Tr\,\rL^2=1$, and in this order in $\epsilon$, the statistical distribution of $Y_{(1)}$ does  not depend on $\psi$ at all.   If the low energy fields are in a mixed state,
  then the distribution  of $Y_{(1)}$ does depend on the state of the low energy fields, but only through $\Tr\,\rL^2$, which measures the entanglement of the low energy fields with some other
  system.  Since in general  $0<\Tr\,\rL^2\leq 1$, the dependence  of $\sqrt{\E[Y_{(1)}^2] }$ on the state of the low energy fields is always bounded by  $\epsilon\sigma_W$ times a factor
  of order 1.  Its dependence on $\chi$ has a similar magnitude.
  
How large is $\epsilon \sigma_W$ compared to the leading term $\epsilon^2\sigma_W^2  \Xi d_2^2$ in the gravitational backreaction? For this, we have to ask what are reasonable values of
$\epsilon$ and $d_2$ in the context of gravity.   In gravity, the backreaction of low energy matter on the geometric entropy $\frac{\sf A}{4G}$ is of order 1, so we want $ \epsilon^2\sigma_W^2\Xi  d_2^2\sim 1$, and hence
 $\epsilon\sigma_W \sim\frac{1}{\sqrt{\Xi}d_2}$.   On the other hand, the
entropy of the short distance modes is of order $\log d_2$.  In gravity, this is $\frac{c}{G}$ for a constant $c$, so $d_2\sim e^{c/G}$ and thus $\epsilon\sigma_W \sim \frac{1}{\sqrt{\Xi} d_2}\sim 
\frac{1}{\sqrt{\Xi}} e^{-c/G}$ is
exponentially small.     Thus this source of  dependence of gravitational backreaction on the state of the low energy fields  is extremely small.
Likewise, although the variance of $Y_{(1)}$ means that the geometric entropy depends on a specific choice of $W$,
this dependence
 is extremely small.  It  is actually comparable to  the dependence of the geometric entropy on 
the very small $W$-dependent fluctuations  of  $D_{A_1A_2}$ around its mean value, and much larger than the dependence on fluctuations in $D_{A_1}$.

 \subsection{Independence from the Recovery}\label{indep}

So far we have not taken into account that in the CCKLP setup, the recovery is supposed to be optimized over the choice of local unitaries $R_A,$ $R_\bA$.
In particular, $R_A$ acts on $\sigma^{(\eps)}_{A_1A_2}$ by conjugation.   Since the optimal $R_A$ will depend on $W$ and $\epsilon$, optimizing over the choice of $R_A$
will modify the $\O(\epsilon^2)$ term in the geometric entropy  relative to what we have calculated so far.   However,
we will argue that this modification is negligible.

It is actually useful to go back to the original formula $ S_\PA(A)=S(\sigma_A)-S(\sigma_{A_1}^{(R)})$ for the geometric entropy (eqn. (\ref{zilcox})).   Here, of course,
$\sigma_{A_1}^{(R)}$ depends on both $W$ and $R$, though we only show $R$ in the notation; for each $W$, we are interested in the effects of optimizing with respect to $R=(R_A,R_{\bA})$.   
At $R=1$, we have  found that the backreaction on $S_\PA(A)$ is of order $\epsilon^2 \Xi d_2^2$.   We want to show that the shift due to optimizing with respect
to $R$ is negligible compared to this -- irrespective of the precise choice of optimization criterion.\footnote{The criterion proposed by CCKLP was described in section 2.
Another plausible criterion would be to minimize $D_{A_1}^{(R)}$.}  Since the action of local unitaries does not affect $S(\sigma_A)$ at all, the only term in $S_\PA(A)$ that
is affected by the optimization is $-S(\sigma_{A_1}^{(R)})$.   For this, we can write a convenient formula
\be\label{inzop} S(\sigma_{A_1}^{(R)})= S(\sigma_{A_1}^{(0)}) - D(\sigma_{A_1}^{(R)}||\sigma_{A_1}^{(0)}) -\Tr\,\delta\sigma_{A_1}^{(R)}\log \sigma_{A_1}^{(0)},\ee
with $\delta\sigma_{A_1}^{(R)}=\sigma_{A_1}^{(R)}-\sigma_{A_1}^{(0)}$.
On the right hand side of eqn. (\ref{inzop}), the first term does not depend on $R_A$ at all.   As for $ D(\sigma_{A_1}^{(R)}||\sigma_{A_1}^{(0)})$, which we will call $D_{A_1}^{(R)}$,
 it is a measure of the error in entanglement wedge reconstruction.
At $R=1$ (but $\epsilon\not=0$), we computed $D(\sigma_{A_1}^{(R)}||\sigma_{A_1}^{(0)})$ and found it to be of order $\epsilon^2\sigma_W^2 \Xi$, tiny compared to the gravitational backreaction of order
$\epsilon^2\sigma_W^2 \Xi  d_2^2$.
The purpose of optimizing over $R_A$ is to reduce the error in entanglement wedge reconstruction, so any reasonable criterion for this
optimization will tend to reduce $D(\sigma_{A_1}^{(R)}||\sigma_{A_1}^{(0)})$ and will have negligible probability to significantly increase it. 
 Since $D(\sigma_{A_1}^{(R)}||\sigma_{A_1}^{(0)})$ is bounded below by zero, the optimization cannot reduce it by more than $\epsilon^2\sigma_W^2 \Xi$.    Combining these statements,  the optimization
changes the second term on the right hand side of eqn. (\ref{inzop}) at most  by a negligible amount of order $\epsilon^2\sigma_W^2\Xi$. 

 It remains only to bound how much the 
optimization affects the last term in eqn. (\ref{inzop}), namely
\be\label{zort} T=-\Tr_{A_1}\delta\sigma_{A_1}^{(R)}\,\log\sigma^{(0)}_{A_1}.
\ee
 For \emph{any} choice
of $R_A$, since $\Tr_{A_1}\delta\sigma_{A_1}^{(R)}=0$,  we can subtract a constant $c$  from
$\log\sigma^{(0)}_{A_1}$ without changing $T=-\Tr_{A_1}\delta\sigma_{A_1}^{(R)} \log\sigma_{A_1}^{(0)}$. The quantum Pinsker inequality asserts that for any two density
matrices $\t\sigma,\sigma$, one has\footnote{For a matrix $X$, one defines $||X||_1=\Tr\,\sqrt{X^\dagger X}$; if $X$ is hermitian,  $||X||_1$ is the sum of the absolute
values of the eigenvalues of $X$.}    $||\t\sigma-\sigma||_1\leq \sqrt{2 D(\t\sigma||\sigma)}$.   Applying this with $\sigma=\sigma_{A_1}^{(0)}$, $\t\sigma=\sigma_{A_1}^{(R)}$,
we get $||\delta \sigma_{A_1}^{(R)}||_1\leq \sqrt{2 D_{A_1}^{(R)}}$. 
To turn the upper bound on $||\delta \sigma_{A_1}^{(R)}||_1$ into an upper bound on $|T|$, we use the fact that for hermitian
 matrices $X,Y$, one has $|\Tr\,XY|\leq ||X||_1 \, ||Y||_\infty$, where $||Y||_\infty$ is the operator norm of $Y$ (the absolute value of its largest eigenvalue). For
a suitable constant $c$, $||\log \sigma_{A_1}^{(0)}-c ||_\infty =\frac{1}{2}\log(p_{\rm{max}}/p_{\rm{min}})$, where $p_{\rm{max}}$ and $p_{\rm{min}}$ are the largest and
smallest eigenvalues of $\sigma_{A_1}^{(0)}$.
Putting these facts together, we get 
\be\label{zeffog} |T|\leq ||\delta\sigma_{A_1}^{(R)}||_1 ||\log\sigma_{A_1}^{(0)}-c||_\infty\leq \sqrt{2 D_{A_1}^{(R)}} \frac{1}{2}\log(p_{\rm{max}}/p_{\rm{min}}).\ee
As already remarked, at $R=1$, $D_{A_1}^{(R)}$ is of order   $\epsilon^2\sigma_W^2 \Xi$, and for the optimal recovery, it will not be significantly bigger. As explained in section \ref{preliminaries},
within the range of validity of the CCKLP model,  $p_{\rm {min}}/p_{\rm{max}}$ will not be exponentially small, so we should view  $\log(p_{\rm{max}}/p_{\rm{min}})$ in eqn.
(\ref{zeffog}) as a number of order 1 and in particular independent of $d_2$.   
The upshot is that eqn. (\ref{zeffog}) bounds $|T|$, and therefore the correction to the geometric entropy because of the recovery,
by an amount of order $\epsilon\sigma_W$, with a coefficient independent of $d_2$.   As explained in section \ref{general}, where a contribution of the same order was found,
this is completely negligible compared to the leading order gravitational backreaction $\epsilon^2\sigma_W^2 \Xi d_2^2$.

   We should note, however, that the bound we have found is likely not optimal.   Quite likely, the correction to the geometric entropy  from the 
recovery procedure is really of order $\epsilon^2\sigma_W^2$ times a coefficient independent of $d_2$.   But to show this would probably involve much more detailed study of the
recovery procedure.

Concerning $D_{A_1}$, as we have already remarked, since it is a measure of the reconstruction error, any sensible optimization over $R_A$ will not significantly increase it.   
So after optimization, the ratio $D_{A_1}/D_{A_1A_2}$ remains tiny.

\subsection{Mixed Bulk State}\label{mixed}

The same methods, with minor changes, apply to the case that the low energy fields in $A_1\bA_1$ are in an arbitrary mixed state $\rL$,
rather than the pure state that we have assumed up to this point.
In this case, at $\epsilon=0$, the state of the full system $A\bar A$ is 
$\rho=\rL\otimes\ket\chi\bra\chi$. 
The only change in eqn. \eqref{MM} is that we cannot set $\rho^2=\rho$.  We will assume that $\chi$ is maximally mixed (this assumption can be relaxed as in section
\ref{general}).
Now let us reconsider the derivation of \eqref{resolved}.  We return to eqn. (\ref{usefulone}) and again use (\ref{MM}) for the second moment of $M$.
   First let us evaluate the contribution from the term  
$-\sigma_W^2\delta_{IL}(\rho^2)_{KJ}$ in eqn. (\ref{MM}).   With the same
relabeling of indices as before ($I=(s\bar s)$, etc.), this becomes $-\delta_{ss}\delta_{\bar s\bar t} (\rho^2)_{(s'\bar t)(s'\bar s)}$.  Summing over $\bar s$ and $\bar t$
to evaluate the contribution to eqn. (\ref{MM}), we get $\epsilon^2\sigma_W^2 (\Tr_{S^c}\,\rho^2)_{s's'}$.   The term in (\ref{MM}) proportional to $\delta_{JK}$
similarly  gives $\epsilon^2\sigma_W^2(\Tr_{S^c}\,\rho^2)_{ss}$.    As for  the $\rho_{KJ}\rho_{IL}$ term in \eqref{MM}, after substituting $I=(s\bar s)$, etc.,  it
becomes $-2\epsilon^2\sigma_W^2 \cE_{ss'}$, with
\be\label{sublabel} \cE_{ss'}=\sum_{\bar s\bar t} \rho_{(s'\bar t)(s'\bar s)}\rho_{(s\bar s)(s\bar t)}.\ee
This quantity has no useful  simplification in general.  So the generalization of (\ref{resolved}) is
\be\label{tublabel} \E\left[|\delta\sigma_S)_{ss'}|^2\right]=\epsilon^2\sigma_W^2\left( (\Tr_{S^c}\,\rho^2)_{ss}+(\Tr_{S^c}\,\rho^2)_{s's'}-2\cE_{ss'}\right). \ee

Now let us evaluate this for $S=A_1$ and $S^c=\bA_1 A_2\bA_2$.    We have $\rho=\rL\otimes \ket{\chi}\bra{\chi}$, and $\rho^2=\rL^2\otimes \ket{\chi}\bra{\chi}$.   So $\Tr_{S^c}\,\rho^2=\Tr_{\bar A_1}\,\rL^2 \cdot \Tr_{A_2\bar A_2}\ket{\chi}\bra{\chi}=\Tr_{\bar A_1}\,\rL^2$.   We also need to evaluate
$\cE_{ss'}$.  For $S=A_1$, indices $s$ and $s'$ label basis vectors $i$ and $j$ of $A_1$, while $\bar s$ and $\bar t$ represent triples, $\bar s=(\mu,m,\alpha),
\,\bar t=(\nu,n,\beta)$, where in each case the three entries of the triple refer respectively to basis vectors of $A_2$, $\bA_1$, and $\bA_2$.   
So $\rho_{(i\bar s)(i\bar t)}$ becomes $\rL_{(im)(in)}(\ket{\chi}\bra{\chi})_{(\mu\alpha)(\nu\beta)}=\rL_{(im)(in)}\frac{\delta_{\mu\alpha}\delta_{\nu\beta}}{d_2}$,
and similarly $\rho_{(j\bar t)(j\bar s)}=\rL_{(jn)(jm)}\frac{\delta_{\mu\alpha}\delta_{\nu\beta}}{d_2}$.   So
\be\label{thelp}\cE_{ij}=\sum_{\mu,m,\alpha}\sum_{\nu,n,\beta}\frac{\delta_{\mu\alpha}\delta_{\nu\beta}}{d_2^2}\rL_{(im)(in)}\rL_{(jn)(jm)}
=\sum_{m,n}\rL_{(im)(in)}\rL_{(jn)(jm)}.\ee
Hence
\be\label{uhelp} \E\left[|(\delta\sigma_{A_1})_{ij}|^2\right]=\epsilon^2\sigma_W^2 \left((\tau_a)_{ii}+(\tau_a)_{jj}-2C_{ij}\right),\ee
where
\be\label{oddmats}\tau_a=\Tr_{\bA_1}\,\rL^2,\qquad C_{ij}=\Tr_{\bA_1}\,\left(\bra{i}\rL\ket{i}\bra{j}\rL\ket{j}\right), \ee
with $\bra{i}\rL\ket{i}$ being the operator on $\bA_1$ whose matrix elements are $\bra{i}\rL\ket{i}_{mn}=\rL_{(im)(in)}$, and similarly for  $\bra{j}\rL\ket{j}$.

The derivation for $S=A_1A_2$ is similar, except that we do not get factors of $d_2$ that came in the last paragraph from summing over indices $\mu,\nu$.
The result is 
\begin{align}\label{vhelp}
\mathbb E\left[|(\delta\sigma_{A_1A_2})_{i\mu,j\nu}|^2\right]
&=\eps^2\sw^2\Big[\frac{(\tau_a)_{ii}+(\tau_a)_{jj}}{d_2}
-\frac{2}{d_2^2}\,\delta_{\mu\nu}\,C_{ij}\Big].
\end{align}
Dropping the assumption that $\rL$ is pure has made it more complicated to describe the numerators in (\ref{uhelp}) or (\ref{vhelp}), but has no effect on 
the powers of $d_2$ that accompany the various terms.

Eqns. (\ref{uhelp}) and (\ref{vhelp}) generalize the previous formulas (\ref{mA1}) and (\ref{mA}) to the case that $\rL$ is not assumed to be pure.
With these generalizations at hand, the rest of the derivation is precisely as before.   We choose the basis vectors $i,j$ of $A_1$ in the preceding formulas
to be eigenvectors of $\sigma_{A_1}^{(0)}$, with corresponding eigenvalues $p_i$.
Proceeding as in section \ref{combining}, everything can be expressed in terms of
\begin{equation}
\Xi_\tau\equiv\sum_{ij}w(p_i,p_j)\,(\tau_a)_{ii},\qquad
\Gamma\equiv\sum_{ij}w(p_i,p_j)\,C_{ij},
\end{equation}
with the result
\begin{equation}\label{ffinal}
\boxed{\;
\mathbb E[D_{A_1}]=\eps^2\sw^2\,(\Xi_\tau-\Gamma),\qquad
\mathbb E[D_{A_1A_2}]=\eps^2\sw^2\,(d_2^2\,\Xi_\tau-\Gamma),\qquad
\frac{\mathbb E[D_{A_1}]}{\mathbb E[D_{A_1A_2}]}\approx
\frac{1}{d_2^2}\Big(1-\frac{\Gamma}{\Xi_\tau}\Big).
\;}
\end{equation}
The pure case is $\Xi_\tau\to\Xi$, $\Gamma\to1$. For mixed $\rL$, the formula is sensitive to the bulk state but only through an  $\O(1)$ prefactor; the $d_2^2$ enhancement of $\E[D_{A_1A_2}]$ over $\E[D_{A_1}]$ comes entirely from the $A_2$ sector and does not depend on the bulk state of the low energy fields. 

As an interesting special case, suppose that the bulk state is maximally mixed, with a density matrix that is
a multiple of the identity.   In this case, $(\tau_a)_{ii}=\frac{1}{d_1^3}$, $C_{ij}=\frac{\delta_{ij}}{d_1^3}$, so
 $\Xi_\tau=\Gamma$, and $\mathbb E[D_{A_1}]=0$ at
this order while $\mathbb E[D_{A_1A_2}]\ne0$.  
The reconstruction error vanishes because a featureless state
has nothing to reconstruct.  
  
\vskip1cm
 \noindent {\it {Acknowledgements}}  
  Research supported in part by NSF Grant PHY-2514611.

 \bibliographystyle{unsrt}

\begin{thebibliography}{99}

\bibitem{Ent1} B. Czech, J. L. Karczmarek, F. Nogueira, and M. Van Raamsdonk,
``The Gravity Dual of a Density Matrix,''  Class.  Quant.  Gravity  {\bf 29} (2012) 155009, arXiv:1204.1330.

\bibitem{Ent2}  M. Headrick, V. E. Hubeny, A. Lawrence, and M. Rangamani, ``Causality \& Holographic Entanglement Entropy,''  JHEP {\bf 12}  (2014) 162,
arXiv:1408.6300.



\bibitem{Ent3}  A. C. Wall,  ``Maximin Surfaces, and the Strong Subadditivity of the Covariant Holographic
Entanglement Entropy,''  Class. Quant.  Gravity {\bf31} (2014) 225007,  arXiv:1211.3494.




\bibitem{Ent4}  D. L. Jafferis, A.  Lewkowycz, J. Maldacena, and S. J. Suh,  ``Relative Entropy
Equals Bulk Relative Entropy,''  JHEP  {\bf 06} (2016) 004, arXiv:1512.06431.

\bibitem{Ent5} X. Dong, D. Harlow, and A. C. Wall, ``Reconstruction of Bulk Operators Within the
Entanglement Wedge in Gauge-Gravity Duality,''  Phys. Rev. Lett. {\bf  117} (2016) 021601, arXiv:1601.05416.


\bibitem{Ent6} J.  Cotler, P. Hayden, G. Penington, G.  Salton, B. Swingle, and
M. Walter,  ``Entanglement Wedge Reconstruction Via Universal Recovery Channels,'' 
 Phys. Rev. {\bf X9} (2019) 031011,
arXiv:1704.05839.




\bibitem{Harlow}D. Harlow, ``The Ryu-Takayanagi Formula From Quantum Error Correction,'' Commun. Math. Phys. {\bf 354} (2017) 865, arXiv:1607.03901.

\bibitem{Kelly}W. R. Kelly, ``Bulk Locality and Entanglement Swapping in AdS/CFT,'' JHEP {\bf 03} (2017) 153, arXiv:1610.00669.

\bibitem{Corr2}P. Hayden and G. Penington, ``Learning the Alpha-bits of Black Holes,'' JHEP {\bf 12} (2019) 007, arXiv:1807.06041.

\bibitem{Corr3} G. Penington, ``Entanglement Wedge Reconstruction and the Information Paradox,''  JHEP {\bf 09} (2020) 002 arXiv:1905.08255.

\bibitem{Corr4} 
P. Faist, S. Nezami, V. V. Albert, G. Salton, F.  Pastawski, P. Hayden, and J. Preskill, ``Continuous Symmetries and Approximate Quantum Error Correction,''
Phys.Rev. {\bf X10} (2020) 041018, arXiv:1902.07714.



\bibitem{Corr5}C.-J. Cao, ``Non-Trivial Area Operators Require Nonlocal Magic,'' JHEP {\bf 11} (2024) 105, arXiv:2012.00199.

\bibitem{CCKLP}C.-J. Cao, G. Cheng, K.Karthikeyan, C. Li, and J. Preskill, ``State-Dependent Geometries From Magic-Enriched Quantum Codes,''
arXiv:2603.13475.

\bibitem{AkersPenington}C. Akers and G, Penington, ``Quantum Minimal Surfaces From Quantum Error Correction,'' SciPost. Phys. {\bf 12}  (2022) 157, arXiv:2109.14618.


\bibitem{LM}A. Lewkowycz and J. Maldacena, ``Generalized Gravitational Entropy,'' JHEP 08 (2013)
090, arXiv:1304.4926.

\bibitem{Der1}T. Barrella, X. Dong, S. A. Hartnoll and V. L. Martin, ``Holographic Entanglement Beyond
Classical Gravity,'' JHEP 09 (2013) 109, arXiv:1306.4682.

\bibitem{Der2}T. Faulkner, A. Lewkowycz and J. Maldacena, ``Quantum Corrections to Holographic
Entanglement Entropy,'' JHEP 11 (2013) 074, arXiv:1307.2892.

\bibitem{ADH}A. Almheiri, X. Dong, and D. Harlow, ``Bulk Locality and Quantum Error Correction in AdS/CFT,'' JHEP {\bf 04} (2015) 163,  arXiv:1411.7041.

\bibitem{HaPPY}F. Pastawski, B. Yoshida, D. Harlow, and J. Preskill, ``Holographic Error-Correcting Codes: Toy Models for the Bulk-Boundary Correspondence,''
JHEP {\bf 06}  (2015) 149, arXiv:1503.06237.

\bibitem{StrohWitten}A. Strohmaier and E. Witten, ``The Timelike Tube Theorem In Curved Spacetime,''
Commun. Math. Phys. {\bf 405} (2024) 7, 153, arXiv:2303.16380. 
\end{thebibliography}

\end{document}